\documentclass[twocolumn,prl,showpacs,amsmath,amssymb]{revtex4-1}
\usepackage{amsmath}
\usepackage{dcolumn}
\usepackage{wrapfig}
\usepackage{graphicx}
\usepackage{color}
\usepackage{graphics}
\usepackage{epsfig}
\usepackage{units}
\usepackage{dsfont}

\makeatletter
\begin{document}
\title{Efficient reconstruction of transmission probabilities in a spreading process from partial observations} 
\author{Andrey Y. Lokhov$^{1,}$}\email{lokhov@lanl.gov}\author{Theodor Misiakiewicz$^{2}$}
\affiliation{$^{1}$Center for Nonlinear Studies and Theoretical Division T-4, Los Alamos National Laboratory, Los Alamos, NM 87545, USA,}
\affiliation{$^{2}$Departement de Physique, Ecole Normale Sup\'erieure, 45 rue d'Ulm, 75005 Paris, France.}
\date{\today}

\begin{abstract}
An important problem of reconstruction of diffusion network and transmission probabilities from the data has attracted a considerable attention in the past several years. A number of recent papers introduced efficient algorithms for the estimation of spreading parameters, based on the maximization of the likelihood of observed cascades, assuming that the full information for all the nodes in the network is available. In this work, we focus on a more realistic and restricted scenario, in which only a partial information on the cascades is available: either the set of activation times for a limited number of nodes, or the states of nodes for a subset of observation times. To tackle this problem, we first introduce a framework based on the maximization of the likelihood of the incomplete diffusion trace. However, we argue that the computation of this incomplete likelihood is a computationally hard problem, and show that a fast and robust reconstruction of transmission probabilities in sparse networks can be achieved with a new algorithm based on recently introduced dynamic message-passing equations for the spreading processes. The suggested approach can be easily generalized to a large class of discrete and continuous dynamic models, as well as to the cases of dynamically-changing networks and noisy information.
\end{abstract}


\maketitle

\paragraph*{Introduction:}

Learning an underlying graphical model from observed data is a long-standing and important practical problem in statistical physics, machine learning and computer science. Recent years have seen a renewed interest in development of fast and efficient algorithms to carry out this reconstruction problem in diverse contexts, such as gene regulatory networks \cite{Marbach2012}, biopolymer structures \cite{Morcos2011a}, neuroscience \cite{Schneidman2006,Cocco2009} and sociology \cite{Eagle2009}, from large datasets becoming available in these fields. An ongoing effort of scientific community has allowed to develop a number of techniques for solving the inverse problem for simple, but widely applicable models, such as the Ising model in static \cite{Roudi2009,Cocco2011,Ravikumar2010,Decelle2014,Bresler2015} and dynamic \cite{Roudi2011,Mezard2011a,Zhang2012,Zeng2013,Bresler2014} settings. However, inference of parameters in a large class of important models of diffusion type has been less studied so far. Among a broad range of disordered and out-of-equilibrium dynamic models, a particular attention is devoted to cascading processes which are used for modelling phenomena in various domains: epidemic and rumor spreading \cite{Hethcote00,Boccaletti2006}, spreading of information and innovations in real-world and virtual social networks \cite{Rogers2010,Strang1998}, avalanches in magnetic and glassy systems \cite{Dhar1997}, activation cascades in neural networks \cite{O'Dea2013}, \emph{etc}.

Contrary to the case of recurrent spreading models, in which the network reconstruction can be achieved via observing one realization of dynamics of sufficient duration \cite{Shen2014,Han2015}, learning in the case unidirectional (also called progressive) dynamics requires a certain number of independent cascades with varying initial conditions. Given a subset of activation times for several realizations of the spreading process, the reconstruction problem aims to infer the transmission parameters of the model. In Bayesian framework, a common inference method relies on the maximization of the likelihood of observed information. In the case of fully observed cascades, this approach has been indeed suggested in a number of recent papers \cite{Myers2010a,Gomez-Rodriguez2011,Du2012,Gomez-Rodriguez2013a,Daneshmand2014}, leading to distributed convex optimization algorithms and outperforming previously suggested heuristics. However, in the majority of realistic applications, it is very difficult or even practically impossible to monitor the state of each and every node over the whole duration of the diffusion process; hence a need to develop reconstruction algorithms which would be able to infer the parameters of the model in the presence of hidden nodes or incomplete time information on the cascades, as well as being robust with respect to the noise in the observations. Despite the importance of this problem, the case of incomplete information in a dynamic process has been poorly addressed so far. The corresponding learning problem for kinetic Ising model has been treated in \cite{Dunn2013a} by means of the path-integral approach and the mean-field methods. In the context of cascading processes, the work \cite{Sefer2015} addressed the network learning problem using relaxation optimization techniques, assuming as an input a full probabilistic trace for each node observed at a subset of times.

In this Letter, we develop a systematic framework for parameter estimation in a spreading model from incomplete observations. As a first natural step in solving this problem, we introduce two algorithms based on maximization of the likelihood of incomplete information via exact marginalization and approximate completion of missing data with the Monte Carlo (MC) sampling, which however appear to be eminently costly; this represents an important limitation of these schemes and makes their use impossible in the applications where a fast online learning is desired. As an alternative which would allow to considerably improve the computation time, we develop a new algorithm based on recently introduced dynamic message-passing (DMP) equations for the spreading processes \cite{Lokhov2015}. These equations allow for an asymptotically exact computation of marginal probabilities of node activation on loopy-but-sparse networks, and can be used as an approximate tool for solving computationally hard problems: recently, DMP equations have been applied to the problem of inference of epidemic origin from a given snapshot of the process at a certain time \cite{Lokhov2014}.

\paragraph*{Formulation of the problem:}

Let $G\equiv(V,E)$ be a connected undirected graph containing $N$ nodes defined by the set of vertices $V$ and the set of edges $E$. We observe $M$ independent realizations of cascades $c$, where each sample $\Sigma^{c}$ represents a set of activation times for the nodes in the network $\{\tau_{i}^{c}\}_{i \in V}$. However, some information on the cascades might be missing: the full information can be written as $\Sigma=\Sigma_{\mathcal{O}}\cup\Sigma_{\mathcal{H}}$, where $\Sigma_{\mathcal{O}}$ is the observed part of the cascades, and $\Sigma_{\mathcal{H}}$ represents the hidden part. For the sake of simplicity and definiteness, we assume that the activation process follows a discrete-time susceptible-infected (SI) model, which is defined as follows \cite{Boccaletti2006}: each node $i$ at time $t \in [0,T]$ can be in one of two states $q_{i}(t)$: susceptible, $q_{i}(t)=S$, or infected, $q_{i}(t)=I$. At each time step, an infected node $j$ can transmit the information to one of its susceptible neighbors $i$ on the interaction graph $G$ with probability $\alpha_{ji}$, meaning that $i$ changes its state with a probability $P_{t}(S(i)\rightarrow I(i))=1-\prod_{k \in \partial i}(1-\alpha_{ki}\mathds{1}[q_{k}(t)=I])$, where $\partial i$ denotes the set of neighbors of $i$; once the node is activated, it stays in the infected state forever. Note that the setting can be straightforwardly generalized to models with more complicated transition rules, such as SIR, threshold and rumor spreading models \cite{Lokhov2015}. We assume that the cascades are simulated with the following initial condition: each node is independently drawn as infected with probability $1/N$, meaning that on average there is one ``patient zero'' at initial time; note, however, that it also means that some cascades have several initial sources, while other cascades are trivial and do not contain any infected nodes. 

Our goal is to reconstruct the values of the couplings $\{\alpha_{ij}\}_{(ij) \in E} \equiv G_{\alpha}$. In the absence of any prior on the underlying model, Bayes' theorem states that
\begin{align}
G_{\alpha} = \arg\max P(G_{\alpha} \mid \Sigma_{\mathcal{O}})
\propto \arg\max P(\Sigma_{\mathcal{O}} \mid {G}_{\alpha}),
\label{eq:general_problem_formulation}
\end{align}
where $\Sigma_{\mathcal{O}} \equiv \{\Sigma^{c}_{\mathcal{O}}\}_{c \in [1,M]}$. Hence, the task is to estimate efficiently the likelihood function $P(\Sigma_{\mathcal{O}} \mid {G}_{\alpha})$. Note that the formulation \eqref{eq:general_problem_formulation} is valid for the case where the structure of the graph is unknown (defining the problem on a fully-connected graph). However, in what follows and unless stated otherwise, we assume that the network $G$ is known; treating the case of unknown graph with missing information would require some additional assumptions and constraints on the network structure. For the tests involving incomplete observations, we focus for definiteness on the presence of nodes with hidden information, providing the study of other cases in the Supplemental Material (SM) \cite{SupplementalInformation}. 

\paragraph*{Maximum likelihood estimator:}

If the information on all the nodes is available ($\Sigma=\Sigma_{\mathcal{O}}$), an efficient strategy would be to use a consistent maximum likelihood estimator (MLE), suggested in \cite{Gomez-Rodriguez2011}. In the discrete formulation, the likelihood of the cascades, $P(\Sigma \mid {G}_{\alpha})$, is given by:
\begin{equation}
P(\Sigma \mid {G}_{\alpha})=\prod_{i \in V} \prod_{1 \leq c \leq M} P_{i}(\tau_{i}^{c} \mid \Sigma^{c} \backslash \tau_{i}^{c}, {G}_{\alpha}),
\label{eq:factorization_likelihood}
\end{equation}
where
\vspace{-0.41cm}
\begin{align}
\notag
&P_{i}(\tau_{i}^{c} \mid \Sigma^{c} \backslash \tau_{i}^{c},  {G}_{\alpha})
=\left(\prod_{t'=0}^{\tau_{i}^{c}-2}\prod_{k \in \partial i}(1-\alpha_{ki}\mathds{1}[\tau_{k}^{c} \leq t'])\right)
\\
&\times\left[1-\left(\prod_{k \in \partial i}(1-\alpha_{ki}\mathds{1}[\tau_{k}^{c} \leq \tau_{i}^{c}-1])\right)\mathds{1}[\tau_{i}^{c}<T] \right].
\label{eq:local_likelihood_discrete}
\end{align}
The estimation of the transmission probabilities ${\widehat{G}_{\alpha}}$ is given by the solution of the convex optimization problem
\begin{equation}
{\widehat{G}_{\alpha}}=\arg\min \left( -\log P(\Sigma \mid {G}_{\alpha}) \right),
\label{eq:MLE_optimization}
\end{equation}
which can be solved locally for each node $i$ and its neighborhood due to the factorization of the likelihood under assumption of asymmetry of the couplings. In the case of partial observations, we need to consider the reduced MLE, performing a trace over the unknown activation times of the hidden nodes:
\begin{equation}
P(\Sigma_{\mathcal{O}} \mid {G}_{\alpha})=\sum_{\{\tau_{i}\}_{i \in \mathcal{H}}}P(\Sigma \mid {G}_{\alpha}).
\label{eq:reduced_likelihood}
\end{equation}
An exact evaluation of \eqref{eq:reduced_likelihood} is a computationally difficult high-dimensional problem with complexity proportional to $T^{H}$, where $H$ is the number of hidden nodes. Unfortunately, we found that this complexity can not be reduced by approximating the objective function \eqref{eq:reduced_likelihood} with an appropriate MC sampling since the evaluation of the gradient should be very precise for convergence of the algorithm; see the SM \cite{SupplementalInformation} for more details.

\paragraph*{Heuristic two-stage algorithm:}

In order to keep the nice convexity properties of the full MLE, we introduce a modification of the scheme above which we also use as a benchmark (this algorithm will be referred to as HTS algorithm). The idea is to use two alternating stages at each step of optimization. First, we complete the missing information in the cascades $\Sigma_{\mathcal{H}}$ using the current estimation of the couplings ${\widehat{G}_{\alpha}}$, i.e. update the activation times of the hidden variables as follows:
\begin{equation}
\widehat{\Sigma}_{\mathcal{H}}=\arg\max P(\Sigma \mid {\widehat{G}_{\alpha}}).
\label{eq:inference_hidden_times}
\end{equation}
Again, an exact solution of \eqref{eq:inference_hidden_times} still requires a number of operations proportional to $T^{H}$, but in practice we approximate the inference problem \eqref{eq:inference_hidden_times} using a MC sampling procedure. Second, we can solve the convex optimization problem \eqref{eq:MLE_optimization} using the ``complete'' $\Sigma=\Sigma_{\mathcal{O}}\cup\widehat{\Sigma}_{\mathcal{H}}$, thus obtaining a new estimation of   ${\widehat{G}_{\alpha}}$; the procedure is repeated until convergence. This leads to an algorithm with complexity $O(\vert E\vert M+NML_{H,T})$ at each step of optimization procedure, where $\vert E\vert$ is the number of edges, and $L_{H,T}$ is a sampling parameter growing with $T$ and $H$ \cite{SupplementalInformation}.

\paragraph*{Dynamic message-passing algorithm:}

A way to quantify the interdependence of activation times of different nodes is to use the dynamic equations that contain information about the correlations occurring in the spreading process. The suggested algorithm is based on the dynamic message-passing equations for diverse dynamic processes \cite{Lokhov2015}. %which allow to compute marginal probabilities of activation of nodes in the network.
According to the DMP equations for the SI model, the marginal probability $m^{i}(\tau_{i})$ of activation of node $i \in V$ at time $\tau_{i}$ can be computed as
\begin{align}
&m^{i}(\tau_{i})=P^{i}_{S}(0)\left[\prod_{k\in \partial i}\theta^{k \rightarrow i}(\tau_{i}-1)\hspace{-0.7mm}-\hspace{-0.9mm}\prod_{k\in \partial i}\theta^{k \rightarrow i}(\tau_{i})\right]
\label{eq:marginals}
\end{align}
for $\tau_{i}>0$, with $m^{i}(0)=1-P^{i}_{S}(0)$, where $P^{i}_{S}(0)$ is the probability that node $i$ is initialized in the state $S$. The quantities $\theta^{k \rightarrow i}(t)$ are computed iteratively using the following expressions:
\begin{align}
& \theta^{k \rightarrow i}(t)=\theta^{k \rightarrow i}(t-1)
-\alpha_{ki}\phi^{k \rightarrow i}(t-1),\label{eq:SIequations:theta}
\\
\notag
&\phi^{k \rightarrow i}(t)=(1-\alpha_{ki})\phi^{k \rightarrow i}(t-1)
\\
&+P_{S}^{k}(0)\prod_{l\in \partial k \backslash i}\theta^{l \rightarrow k}(t-1)-P_{S}^{k}(0)\prod_{l\in \partial k \backslash i}\theta^{l \rightarrow k}(t),
\label{eq:SIequations:phi}
\end{align}
with the initial conditions $\theta^{i \rightarrow j}(0)=1$ and $\phi^{i \rightarrow j}(0)=1-P_{S}^{i}(0)$, and $\partial k \backslash i$ denoting the set of neighbors of $k$ excluding $i$. The proof that these equations are exact on trees and empirical studies of performance on random and real-world networks are discussed in \cite{Lokhov2015}.

Let us now explain the reconstruction algorithm based on the DMP equations. Given the data on the cascades $\Sigma_{\mathcal{O}}$, we can compute the empirical initial conditions $P^{i}_{S}(0)$ and marginal probabilities $m^{i}_{*}(\tau_{i})$, simply given by the averages of activation times over different cascades at all nodes for which the information is known. The idea, reminiscent of what has been previously used in online learning of parameters in the context of artificial neural networks \cite{Williams1989}, is to adjust the transmission probabilities $G_{\alpha}$ in order to minimize the mismatch $J$ between the DMP-estimated and available empirical marginals at each time step:
\begin{equation}
J=\sum_{t=0}^{T-1}J(t)=\sum_{t=0}^{T-1}\sum_{i \in \mathcal{O}} \frac{1}{2} [m^{i}_{*}(t)-m^{i}(t)]^{2}.
\label{eq:mismatch}
\end{equation}

To this end, we use simple gradient descent: starting from some initial distribution of transmission probabilities, the couplings are updated as $\alpha^{(t+1)}_{rs} \leftarrow \alpha^{(t)}_{rs} - \epsilon \frac{\partial J(t)}{\partial \alpha_{rs}}$, where $\epsilon$ is the learning rate. The derivatives of the cost function \eqref{eq:mismatch} with respect to couplings can be expressed through $\frac{\partial \theta^{k \rightarrow i}(t)}{\partial \alpha_{rs}}\equiv p^{k \rightarrow i}_{rs}(t)$ and $\frac{\partial \phi^{k \rightarrow i}(t)}{\partial \alpha_{rs}}\equiv q^{k \rightarrow i}_{rs}(t)$, for which the DMP-like equations can be easily written using an explicit derivation of the equations \eqref{eq:SIequations:theta}-\eqref{eq:SIequations:phi} \cite{SupplementalInformation}. The update of the transmission probabilities is restarted from time zero until the convergence of the algorithm. Because of the averaging over the cascades, the resulting computational complexity of an iteration step of the DMP algorithm is independent on $M$ and $H$ and is equal to $O(N d T)$, where $d$ is the average degree of the graph.

\paragraph*{Performance of reconstruction algorithms:}

\begin{figure}[!b]
\begin{center}
\includegraphics[width=\columnwidth]{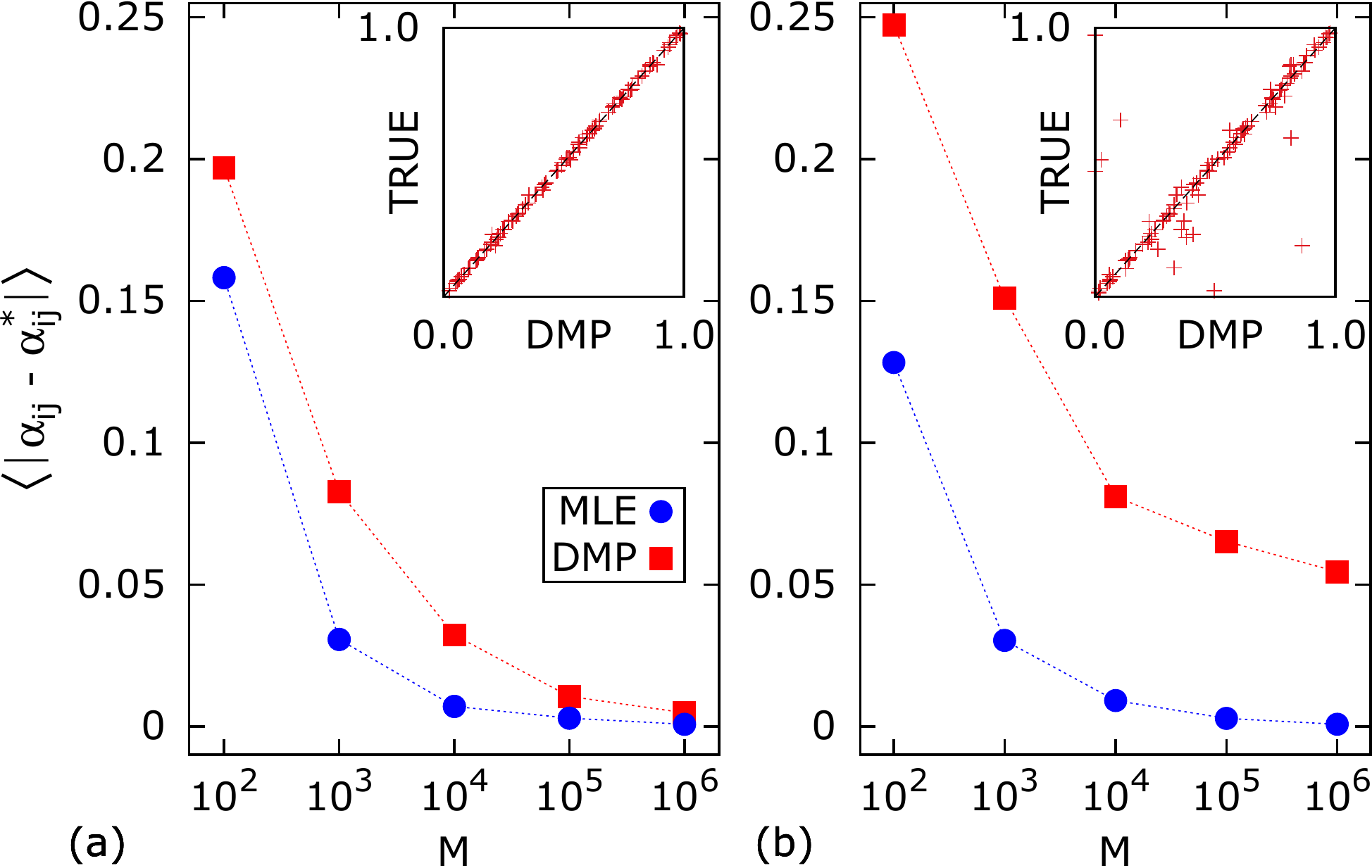}
\caption{(Color online) Main figures: Comparison of mean error on the reconstructed couplings by MLE and DMP algorithms initialized at $\alpha_{ij}=0.5$ for all edges $(i,j) \in E$ as a function of the number of fully observed cascades $M$ for (a) a tree network with $N=50$ and (b) a connected component of a power-law network with $N=53$. Insets: Scatter plots of transmission probabilities reconstructed by DMP algorithm for $M=10^{6}$ versus true couplings for (a) the tree and (b) the power-law network.}
\label{fig:full}
\end{center}
\vspace{-1.56cm}
\end{figure}

\begin{figure}[!ht]
\begin{center}
\includegraphics[width=\columnwidth]{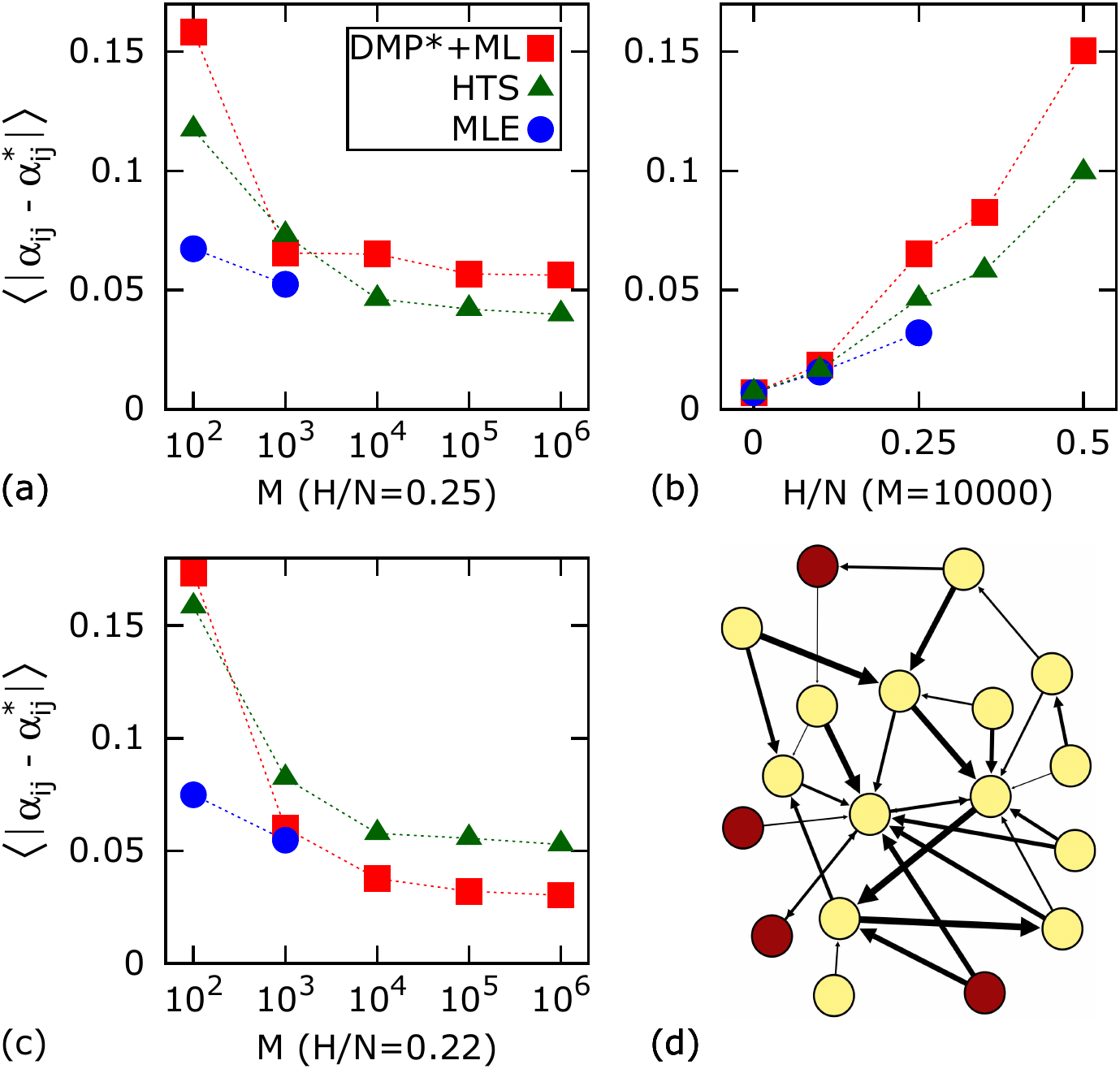}
\caption{(Color online) Comparison of reconstruction performance in terms of the mean reconstruction error for different values of $M$ and $H$ of MLE, DMP*+ML and HTS algorithms on (a), (b) a connected component of a power-law network with $N=20$ and (c) a real-world Sampson's monastery network with $N=18$ nodes \cite{sampson1969crisis,Pajek}; the topology of the latter is visualized in (d) using Gephi \cite{bastian2009gephi}, with darker color indicating the randomly chosen location of nodes with hidden information used in the plot (c).}
\label{fig:incomplete} 
\end{center}
\vspace{-0.7cm}
\end{figure}

We test all the aforementioned algorithms on synthetic and real-world networks with the data generated from $\{\alpha_{ij}\}_{(ij) \in E}$ uniformly distributed in the range $[0,1]$, and using $T=10$. We first illustrate their performance in the case of fully observed cascades. In the Fig.~\ref{fig:full}, we present results for the mean error of reconstruction per coupling on a tree network and on a connected component of an artificially-generated random graph with the Pareto power-law degree distribution of shape parameter $2.5$ and minimum value parameter $1$. As expected, MLE gives a much better prediction of couplings overall compared to the DMP algorithm, because of the loss of information due to the averaging over the statistics of cascades in the latter. Yet, in the case of imperfect measurements, this averaging may play a positive role for the quality of reconstruction: in the SM \cite{SupplementalInformation}, we show that already in the presence of weak noise a naive application of MLE leads to a quite poor reconstruction since this scheme crucially depends on the detailed information contained in the data, while DMP remains much more robust to the noise.
Still, as demonstrated in the insets of the Fig.~\ref{fig:full}, assuming the perfect measurements we see that whereas on a tree network the DMP algorithm yields almost a perfect reconstruction of couplings for a large number of observed cascades, it makes clearly wrong prediction of couplings in the vicinity of very short loops present in the small power-law network due to the inaccuracy of the DMP equations on graphs with small loops and to the degeneracy of solutions leading to the same values of marginal probabilities.

In order to reduce the error of the DMP equations due to short loops which are present in most of realistic networks, we suggest a modification of the algorithm (which we denote DMP* in what follows) based on the following observation: if one chooses to run the DMP equations for only $T' \lesssim l$ time steps, where $l$ is the length of the shortest cycle in the network, and keep only $T'$ terms in the sum \eqref{eq:mismatch}, the marginals predicted by DMP will be exact since the spreading will effectively stay on a tree. The drawback of this choice consists in a further increase of the degeneracy of possible solutions, especially in the presence of hidden nodes; but this feature is proper to all considered schemes since the objective function landscapes naturally develop local minima for loopy graphs. Thus aiming at reinforcing the convergence of algorithms towards the true solution, we choose to initialize them using the following idea: since MLE is still supposed to give a good estimation of the parameters for visible nodes if one ignores the hidden nodes in its neighborhood \cite{SupplementalInformation}, we estimate and ``freeze'' this part of couplings using the fast local maximization of \eqref{eq:factorization_likelihood}, and optimize over all remaining couplings setting them equal to $0.5$ at initial time. If this initialization strategy is used in the DMP algorithm, we denote the corresponding modification by DMP+ML.    

The Fig.~\ref{fig:incomplete} is devoted to the tests in the presence of nodes with hidden information. Because of the large convergence times for MLE and HTS in the case of incomplete information, we were forced to perform tests on small and loopy networks. Note that if the hidden node is a leaf of the network, then no algorithm can reconstruct the transmission probability associated with the ingoing directed edge adjacent to this nodes; therefore, we do not include such parameters in the computation of the mean error throughout the Fig.~\ref{fig:incomplete}. As expected, MLE demonstrates the best performance, but it is not practically applicable even for small networks of size $N \lesssim 20$, and we show only the first few values as reference points for other algorithms. Although in most of the cases the DMP*+ML algorithm is less precise then the HTS approach, overall it demonstrates similar reconstruction results, sometimes even outperforming HTS, as shown in the Fig.~\ref{fig:incomplete}~(c) for a small real-world network of friendship relationships extracted from an ethnographic study of community structure in a New England monastery \cite{sampson1969crisis,Pajek}, depicted in the Fig.~\ref{fig:incomplete}~(d). The whole difference lies in the comparison of computation times: for instance, computation of the point corresponding to $H=4$, $M=10^4$ on a power-law network with $N=20$ in Fig.~\ref{fig:incomplete}~(b) took a time of order of two weeks for MLE, one day for HTS and less than 30 seconds for the DMP algorithm. Basically, it means that the HTS algorithm is not applicable even for reasonably small networks, such as a network of retweets with $N=96$ nodes \cite{graphrepository2013}. Test of the DMP algorithm on this and other real-world networks \cite{Pajek,read1954cultures}, as well as other examples involving incomplete observations in time and noisy information are provided in the SM \cite{SupplementalInformation}.

\paragraph*{Conclusion and perspectives:}

An approximate solution given by the DMP equations for spreading processes allowed us to introduce an efficient algorithm for the reconstruction of transmission probabilities in the presence of hidden information, demonstrating comparable results with respect to the methods based on the maximization of incomplete likelihood at a substantially lower computational cost. It can be used for large networks, providing the best performance in the case of sparse networks. Let us indicate some possible generalizations and perspectives. It would be useful to understand whether the performance of the DMP algorithm could be further improved by matching the two-point correlations. Along with applications to a range of other spreading models \cite{Lokhov2015}, the DMP algorithm can be straightforwardly generalized to the case of continuous-time models using the corresponding version of the DMP equations \cite{KarrerNewman2010} and to dynamically-varying networks in the spirit of \cite{Gomez-Rodriguez2013a} using the time-dependent couplings $\alpha_{ij}(t)$. An interesting future direction would be to adapt the DMP approach for the network structure learning under some restrictions on the class of networks one tries to reconstruct, e.g. using a $\ell_{1}$-regularization \cite{Myers2010a,Daneshmand2014}. Finally, in the spirit of active learning, it might be advantageous to optimize over the particular choice of a reduced number of costly measurements sufficient for an accurate reconstruction. Some of these directions are further discussed in the SM \cite{SupplementalInformation}.

\begin{acknowledgments}
The authors are grateful to M. Chertkov, S. Misra and M. Vuffray for fruitful discussions and valuable comments, as well as to D. Saad and L. Zdeborov\'a for useful references. The work at LANL was carried out under the auspices of the National Nuclear Security Administration of the U.S. Department of Energy under Contract No. DE-AC52-06NA25396. We acknowledge partial support and hospitality of CNLS at LANL. T.M. also acknowledges partial support of the NSF/ECCS collaborative research project on Power Grid Spectroscopy through NMC. The C implementation of the DMP algorithm is available at this url \cite{ProgramLink}.
\end{acknowledgments}

\bibliography{DynamicLearning}

\newpage

\onecolumngrid
\newpage
\appendix
\pagenumbering{arabic}
\renewcommand*{\thepage}{SM \arabic{page}}

\begin{center}
{\Large Supplemental Material}
\linebreak

{\large \textbf{Efficient reconstruction of transmission probabilities in a spreading process from partial observations}}
\linebreak

Andrey Y. Lokhov and Theodor Misiakiewicz
\linebreak
\end{center}

{\small In the Supplemental Material, we provide implementation details for all the algorithms presented in the main text, present some supportive plots and results, as well as discuss some possible future directions outlined in the main text.} 

\section*{Details on the implementation of the algorithms}

\subsection{Exact marginalisation for the MLE}

The estimation of model parameters in the MLE algorithm is based on the maximization of the log-likelihood of observed part of the cascades $\Sigma_{\mathcal{O}} \equiv \{\Sigma^{c}_{\mathcal{O}}\}_{c \in [1,M]}$:
\begin{equation}
\mathcal{L}(\Sigma_{\mathcal{O}}\mid {G}_{\alpha} ) =  \frac{1}{M} \sum_{c=1}^{M} \log P(\Sigma^{c}_{\mathcal{O}} \mid {G}_{\alpha}),  
\label{eq:objfunc}
\end{equation}
where $P(\Sigma^{c}_{\mathcal{O}} \mid {G}_{\alpha})$ in the case of nodes with hidden information represents the likelihood of the observed part of a cascade $c$:
\begin{equation}
P(\Sigma^{c}_{\mathcal{O}} \mid {G}_{\alpha})=\sum_{\{\tau^{c}_{i}\}_{i \in \mathcal{H}}}P(\{ \tau^{c}_i \}_{i \in \mathcal{O}}, \{\tau^{c}_{i}\}_{i \in \mathcal{H}} \mid {G}_{\alpha}).
\label{eq:marginalized}
\end{equation}
The marginalization over unknown activation times in \eqref{eq:marginalized} is a hard high-dimensional integration problem, with an exponential complexity $T^{H}$. A usual approximation scheme used in this kind of problems consists in evaluating the sum \eqref{eq:marginalized} using a biased Monte Carlo sampling. We have encountered the following problem on this way: the objective function is dominated by an ensemble of cascades with small likelihood, and since we need to evaluate \eqref{eq:objfunc} at each step of the gradient descent using the current estimation point of couplings (which may be very distinct from the true ones), we found that in practice the algorithm has convergence issues if \eqref{eq:objfunc} and the gradient
\begin{equation}
\nabla \mathcal{L}(\{\Sigma^{c}_{\mathcal{O}}\}_{c \in [1,M]}\mid {G}_{\alpha} ) =  \frac{1}{M} \sum_{c=1}^{M} \frac{\nabla P(\Sigma^{c}_{\mathcal{O}} \mid {G}_{\alpha})  }{P(\Sigma^{c}_{\mathcal{O}} \mid {G}_{\alpha}) }
\label{eq:gradient_MLE}
\end{equation}
are evaluated approximately, even with a large sampling parameter. Hence, we have used an exact evaluation of the integrals for the MLE algorithm throughout this work, also having in mind the purpose to produce reference points for comparison of the accuracy for other suggested algorithms with a lower computational complexity. Since one needs to evaluate \eqref{eq:objfunc} and \eqref{eq:gradient_MLE} for each of $M$ independent cascades, the resulting computation complexity of the MLE algorithm for each step of the gradient descent is $O(N T^H M)$. In order to speed up the convergence, we first performed a rough gradient descent using a renormalized gradient for a fixed number of steps to get relatively quickly to the proximity of the minimum, and only then used the exact values of the gradient \eqref{eq:gradient_MLE} until the convergence of the algorithm. As a criterion for convergence, we required the variation of the objective function \eqref{eq:objfunc} in two successive steps to be smaller then a certain threshold $\delta_{MLE}$; the value $\delta_{MLE}=10^{-6}$ was used in our simulations.

\subsection{Completion of cascades in the HTS algorithm}

Although we were not able to use approximation schemes for the MLE scheme, the Monte Carlo sampling appears to be an important element of the heuristic two-stage algorithm. In the first stage of the algorithm, we attempt to complete the missing activation times in each cascade $c$ with their most probable values given the current estimation of couplings $\widehat{G}_{\alpha}$, i.e. to solve the following optimization problem:
\begin{equation}
\{\widehat{\tau^{c}_h}\}_{h \in \mathcal{H}}=\arg\max P(\{\tau^{c}_{i} \}_{i \in \mathcal{O}}, \{\tau^{c}_{h} \}_{h \in \mathcal{H}} \mid {\widehat{G}_{\alpha}}).
\label{eq:likelihood_HTS}
\end{equation}
Again, the search space for the missing activation times is in general of the size $T^{H}$, but we significantly reduce the computational time by exploiting the following idea: the likelihood is non-zero only for the activation times which form a possible cascade on the graph $\widehat{G}_{\alpha}$. Hence, for each cascade $c$, we sample $L_{H,T}$ auxiliary cascades starting from the source points in the visible part of the cascade $\{ i \mid i \in  \mathcal{O}, \tau^{c}_{i} = 0 \}$ (if there are any) and from random sources in the hidden part $\mathcal{H}$ selected with the probability $1/N$. We then choose the subset of activation times for the nodes in the hidden part $\mathcal{H}$ (from one of the auxiliary cascades) which maximises the right hand-side of \eqref{eq:likelihood_HTS} as a solution of the first step of the algorithm.

The value of the sampling parameter $L_{H,T}$ should be small enough to improve the speed of the algorithm, and large enough to allow for a precise inference of parameters and to lead to a successful convergence. Intuitively, it is clear that $L_{H,T}$ should grow as a $H$ and $T$ increase, but it is hard to get an exact dependency. In practice, we found that a good convergence of the algorithm is ensured for the studied networks of size $N\sim 20$ and $T=10$ with the following values: $L_{H,T} = 100$ for $H\leq 3$ , $L_{H,T}=1000$ for $4 \leq H\leq 6$ and $L_{H,T}=10000$ for $7 \leq H \leq 10$. The computational complexity of completion of $M$ cascades using the MC sampling is $O(N M L_{H,T})$. In the second stage of the algorithm, we need to preprocess the hence ``complete'' data of the cascades for the solution of the convex optimization problem (4) of the main text, and this requires $O(\vert E\vert M)$ of operations. The two stages are alternated until the global convergence of the algorithm, which we establish as a variation of two consecutive estimations for the transmission probabilities $\{\alpha_{ij}\}_{(ij) \in E}$ is less then the threshold $\delta_{HTS}$. With the values of $L_{H,T}$ given above, we used $\delta_{HTS}=10^{-2}$ in all the simulations. 

\subsection{Computation of the gradient in the DMP algorithm}

In this section, we will provide details for the derivation of the dynamic message-passing equations that we use to compute the gradient of the cost function
\begin{equation}
J(t)=\sum_{i \in \mathcal{O}} \frac{1}{2} [m^{i}_{*}(t)-m^{i}(t)]^{2}
\label{eq:SMcostfunction}
\vspace{-0.1cm}
\end{equation}
at each iteration step of the DMP algorithm:
\begin{equation}
-\frac{\partial J(t)}{\partial \alpha_{rs}}=\sum_{i \in \mathcal{O}}[m^{i}_{*}(t)-m^{i}(t)]\frac{\partial m^{i}(t)}{\partial \alpha_{rs}}.
\label{eq:gradient_DMP}
\end{equation}
For convenience, let us recapitulate the DMP equations:
\begin{align}
&m^{i}(t)=P^{i}_{S}(0)\left[\prod_{k\in \partial i}\theta^{k \rightarrow i}(t-1)\hspace{-0.7mm}-\hspace{-0.9mm}\prod_{k\in \partial i}\theta^{k \rightarrow i}(t)\right],
\label{eq:SMmarginal}
\\
& \theta^{k \rightarrow i}(t)=\theta^{k \rightarrow i}(t-1)
-\alpha_{ki}\phi^{k \rightarrow i}(t-1),\label{eq:SMtheta}
\\
&\phi^{k \rightarrow i}(t)=(1-\alpha_{ki})\phi^{k \rightarrow i}(t-1)
+P_{S}^{k}(0)\prod_{l\in \partial k \backslash i}\theta^{l \rightarrow k}(t-1)-P_{S}^{k}(0)\prod_{l\in \partial k \backslash i}\theta^{l \rightarrow k}(t).\label{eq:SMphi}
\end{align}
Using the equation \eqref{eq:SMtheta}, we have
\begin{equation}
\frac{\partial m^{i}(t)}{\partial \alpha_{rs}}=P^{i}_{S}(0)\left[\sum_{k\in \partial i}\frac{\partial \theta^{k \rightarrow i}(t-1)}{\partial \alpha_{rs}}\prod_{l\in \partial i \backslash k}\theta^{l \rightarrow i}(t-1)-\sum_{k\in \partial i}\frac{\partial \theta^{k \rightarrow i}(t)}{\partial \alpha_{rs}}\prod_{l\in \partial i \backslash k}\theta^{l \rightarrow i}(t)\right].
\label{eq:derivative_marginal}
\end{equation}
Let us introduce useful notations:
\begin{equation}
\frac{\partial \theta^{k \rightarrow i}(t)}{\partial \alpha_{rs}}\equiv p^{k \rightarrow i}_{rs}(t),
\quad \quad \quad
\frac{\partial \phi^{k \rightarrow i}(t)}{\partial \alpha_{rs}}\equiv q^{k \rightarrow i}_{rs}(t).
\end{equation}
Since the initial dynamic messages $\{\theta^{i \rightarrow j}(0)\}_{(ij) \in E}$ and $\{\phi^{i \rightarrow j}(0)\}_{(ij) \in E}$ are independent on the couplings, we have $p^{k \rightarrow i}_{rs}(0)=q^{k \rightarrow i}_{rs}(0)=0$ for all $k$, $i$, $r$ and $s$, and these quantities can be computed iteratively using the analogues of the equations \eqref{eq:SMtheta} and \eqref{eq:SMphi} of the main text:
\begin{align}
& p^{k \rightarrow i}_{rs}(t)=p^{k \rightarrow i}_{rs}(t-1)
-\alpha_{ki}q^{k \rightarrow i}_{rs}(t-1)-\phi^{k \rightarrow i}(t-1)\mathds{1}[k=r,i=s],\label{eq:SIequations:p}
\\
\notag
&q^{k \rightarrow i}_{rs}(t)=(1-\alpha_{ki})q^{k \rightarrow i}_{rs}(t-1)-\phi^{k \rightarrow i}(t-1)\mathds{1}[k=r,i=s]
\\
&+P_{S}^{k}(0)\sum_{l\in \partial k \backslash i}p^{l \rightarrow k}_{rs}(t-1)\prod_{n\in \partial k \backslash \{i,l\}}\theta^{n \rightarrow k}(t-1)-P_{S}^{k}(0)\sum_{l\in k \backslash i}p^{l \rightarrow k}_{rs}(t)\prod_{n\in \partial k \backslash \{i,l\}}\theta^{n \rightarrow k}(t).
\label{eq:SIequations:q}
\end{align}
Hence, at each time step, we compute the marginals using the DMP equations \eqref{eq:SMmarginal}-\eqref{eq:SMphi}, and use \eqref{eq:gradient_DMP} and \eqref{eq:derivative_marginal}-\eqref{eq:SIequations:q} to update the couplings. In principle, at each iteration step of the algorithm we could run the DMP equations for all $T$ time steps with the current estimation of the couplings, and only then update the transmission probabilities using the derivative of the total cost function $J=\sum_{t=0}^{T-1}J(t)$; we found that the ``online''-like update at each time step in the spirit of \cite{Williams1989} leads to a faster convergence of the algorithm. An intuition behind this choice is as follows: instead of accumulating the error due to the current estimation of the couplings through the whole process, we adjust the couplings progressively as the process spreads through the network.

In practice, we observed that simple gradient descent with a fixed learning rate $\epsilon$ demonstrated good convergence to the optimum, and this is the procedure that we used for producing all the plots in this paper. For most of the plots, we used $\epsilon=5.0$, and the tolerance on the change of the objective function $\delta_{DMP}=10^{-12}$ as a stopping criterion for the algorithm. We saw that the reconstruction results can be improved with a modification of the algorithm DMP* using a reduced number of time steps $T'$ of order of the length of the shortest cycle in the network for an increased accuracy of the DMP equations.  It would be interesting to see if the convergence properties of the algorithm in very hard cases could be further improved by exploring two straightforward extensions:

\begin{enumerate}
\item Adding to the cost function $J$ terms that would reinforce a matching of the two-point correlations corresponding to the probabilities of observing mean probabilities of pairs $\langle S_{i}(t)S_{j}(t) \rangle$, $\langle S_{i}(t)I_{j}(t)\rangle$ and $\langle I_{i}(t)I_{j}(t)\rangle$ for $(ij) \in E$. In fact, these two-point correlations at equal times can be computed within the DMP approach: the basic relation is $\langle S_{i}(t)S_{j}(t)\rangle = P_{S}^{i \rightarrow j}(t)P_{S}^{j \rightarrow i}(t)$, where
\begin{equation}
P_{S}^{i \rightarrow j}(t) = P^{i}_{S}(0)\prod_{k\in \partial i \backslash j}\theta^{k \rightarrow i}(t)
\end{equation}
has a meaning of the probability that the node $i$ is in the state $S$ at time $t$ in an auxiliary cavity dynamics $D_{j}$, in which the node $j$ is fixed to the state $S$ for all times; see \cite{Lokhov2015} for more details. Once $\langle S_{i}(t)S_{j}(t)\rangle$ are computed, we immediately obtain
\begin{equation}
\langle S_{i}(t)I_{j}(t)\rangle = P_{S}^{i}(t) - \langle S_{i}(t)S_{j}(t)\rangle, \quad \quad
\langle I_{i}(t)I_{j}(t)\rangle = 1-P_{S}^{i}(t) - \langle I_{i}(t)S_{j}(t)\rangle.
\label{eq:correlations}
\end{equation}
Since correlations in \eqref{eq:correlations} can be expressed from the single-node marginals and correlations $\langle S_{i}(t)S_{j}(t)\rangle$, we may try to reinforce only the $S-S$ correlations. We have performed several tests using the following modified objective function:
\begin{equation}
J'(t) = J(t) + \frac{1}{\vert E\vert} \sum_{(ij) \in E_{\mathcal{O}}} \frac{1}{2} \left[\Delta\langle S_{i}(t)S_{j}(t)\rangle_{*} - \Delta\langle S_{i}(t)S_{j}(t)\rangle\right]^{2},
\label{eq:correlations_included} 
\end{equation}
where $E_{\mathcal{O}}$ represents the ensemble of edges between observed nodes in $G$, and $\Delta\langle S_{i}(t)S_{j}(t)\rangle_{*} = \langle S_{i}(t+1)S_{j}(t+1)\rangle_{*} - \langle S_{i}(t)S_{j}(t)\rangle_{*}$, where ``$*$'' denotes empirical correlations obtained from averaging over the given data. The gradient of $J'(t)$ can be easily written, since the derivative of a DMP-computed correlation reads
\begin{equation}
\frac{\partial \langle S_{i}(t)S_{j}(t)\rangle}{\partial \alpha_{rs}} = \frac{\partial \left(P_{S}^{i \rightarrow j}(t)P_{S}^{j \rightarrow i}(t)\right)}{\partial \alpha_{rs}} = \left(P^{i}_{S}(0)\right)^{2} \left( \sum_{k \in i \backslash j} \frac{p^{k \rightarrow i}_{rs}(t)}{\theta^{k \rightarrow i}(t)} + \sum_{l \in j \backslash i} \frac{p^{l \rightarrow j}_{rs}(t)}{\theta^{l \rightarrow j}(t)} \right) \langle S_{i}(t)S_{j}(t)\rangle.
\end{equation}
Our preliminary tests are rather inconclusive: we found that while in general the minimization of \eqref{eq:correlations_included} indeed improves the convergence speed and the accuracy of solution in the case of sparse networks and small number of nodes with missing information, it may degrade the quality of reconstruction in the case of loopy networks and a large number of hidden nodes with respect to the use of $J(t)$, presumably because the inclusion of correlations amplifies the error of the DMP equations in these cases. Further investigations are needed to get a clear understanding under what circumstances the use of correlations can be advantageous.

\item Using the information contained in the second derivatives of the cost function $J$ for a better control over the convergence. Indeed, the second derivatives can be computed in the same message-passing way as the gradient; it would, however, involve manipulations with the sixth-order tensors, as opposed to the fourth-order quantities used in the computation of the gradient \eqref{eq:SIequations:p} and \eqref{eq:SIequations:q}. For example, if we denote $\frac{\partial^{2} \theta^{k \rightarrow i}(t)}{\partial \alpha_{rs}\partial \alpha_{uv}}\equiv p^{k \rightarrow i}_{rs,uv}(t)$ and $\frac{\partial^{2} \phi^{k \rightarrow i}(t)}{\partial \alpha_{rs}\partial \alpha_{uv}}\equiv q^{k \rightarrow i}_{rs,uv}(t)$, and since from \eqref{eq:SMmarginal} we have $m^{i}(t)=P^{i}_{S}(t-1)-P^{i}_{S}(t)$ for $t>0$, where
\begin{equation}
P_{S}^{i}(t) = P^{i}_{S}(0)\prod_{k\in \partial i}\theta^{k \rightarrow i}(t),
\label{eq:SIequations:P_S}
\end{equation}
it is sufficient to compute
\begin{equation}
\frac{\partial^{2} P_{S}^{i}(t)}{\partial \alpha_{rs}\partial \alpha_{uv}} = P_{S}^{i}(0) \sum_{k \in \partial i}\left[ p^{k \rightarrow i}_{rs,uv}(t) \prod_{l\in \partial i \backslash k}\theta^{l \rightarrow i}(t) + p^{k \rightarrow i}_{uv}(t) \sum_{m \in i \backslash k} p^{m \rightarrow i}_{rs}(t) \prod_{l\in \partial i \backslash \{k,m\}}\theta^{l \rightarrow i}(t) \right],
\end{equation}
with $p^{k \rightarrow i}_{rs,uv}(t)$ following dynamic message-passing equations obtained as a derivative of \eqref{eq:SIequations:p} and \eqref{eq:SIequations:q} with respect to $\alpha_{uv}$. Note that $p^{k \rightarrow i}_{rs,uv}(t)$ is indeed a six-tensor, which makes it use less practical; we did not use it in our simulations since the observed convergence was already good enough.
\end{enumerate}

\section*{Additional supportive plots and results}

\subsection{Reconstruction of transmission probabilities in the case of noisy information}

\begin{figure*}[!ht]
\begin{center}
\includegraphics[width=0.5\columnwidth]{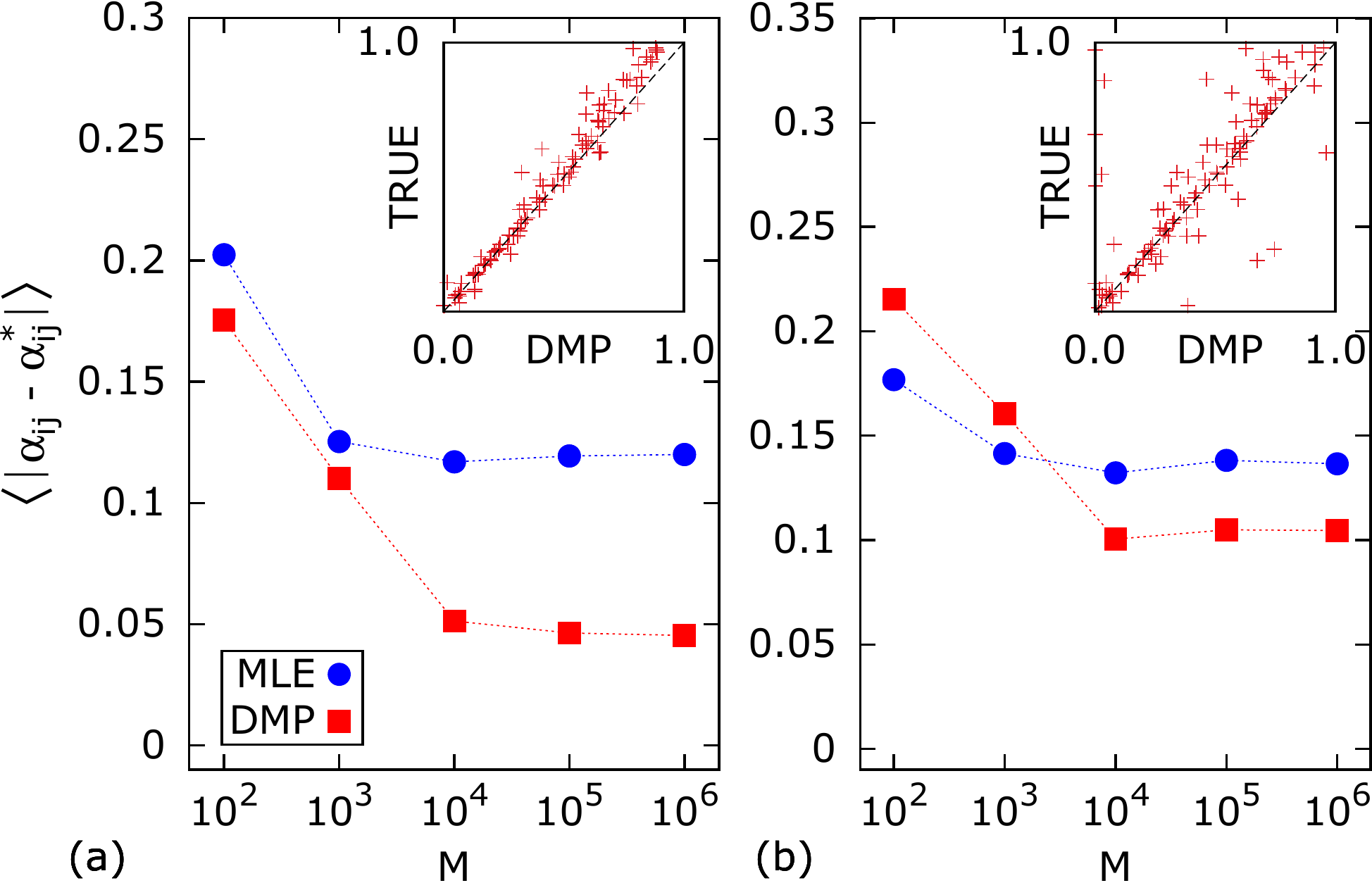}\caption{(Color online) Main figures: Comparison of mean error on the reconstructed couplings by MLE and DMP algorithms initialized at $\alpha_{ij}=0.5$ for all edges $(i,j) \in E$ as a function of the number of fully observed cascades $M$ in the presence of a weak Poissonian noise for (a) a tree network with $N=50$ and (b) a connected component of a power-law network with $N=53$. Insets: Scatter plots of transmission probabilities reconstructed by DMP algorithm for $M=10^{6}$ versus true couplings for (a) a tree and (b) a power-law network.}
\label{fig:Noise}
\end{center}
\vspace{-5mm}
\end{figure*}

In this section, we show that the DMP algorithm is naturally adapted for an efficient reconstruction in the case where the activation times are observed with some noise fluctuating around the true values of the activation time. The reason for that lies in the averaged origin of the empirical marginal probabilities used as an input for the DMP algorithm: while this averaging of the original data may represent a certain drawback since some detailed information on the cascades is lost, in the case of the observations corrupted by noise this procedure has a clear advantage because of the effective averaging over the fluctuations. As a simple test, we have perturbed all the observed activation times $\{\tau_{i}^{c}\}_{i \in V}$ for $c=1\ldots M$ cascades uniformly with a noise $\{\Delta \tau_{i}^{c}\}_{i \in V}$ of randomly chosen sign, with absolute value distributed according to the Poisson distribution with mean $\mu=0.1$. The results of a naive application of MLE and DMP algorithms are presented in the Fig.~\ref{fig:Noise}. We see that since DMP algorithm starts to yield better reconstruction results with respect to the MLE which uses a detailed information on the cascades and performs better in the noiseless case. This relative robustness to noise represents a useful property of the DMP algorithm.

\subsection{Reconstruction of transmission probabilities in the case of incomplete information in time}

Let us discuss another possible type of missing information, when we do not have access to the state of the network at all times, but record its state only at a certain set of observation times $\mathcal{T}_{\mathcal{O}} = \{t^{k}\}_{k=1,\ldots,K}$. For simplicity and without loss of generality, let us assume that the state of the network is recorded at time intervals of equal duration $\Delta$, meaning that $\forall k$, $t^{k+1}-t^{k}=\Delta$. Basically, for each cascade $c$ and node $i$ we don't know the exact value of its activation time $\tau^{c}_{i}$, but only know that it falls inside an interval $\left] (k^{c}_{i}-1) \Delta, k^{c}_{i} \Delta \right]$. A special value $\Delta=1$ corresponds to the case of full information. Once again, MLE requires an exponential number of evaluations proportional to $\Delta^{N}$, so we do not consider it here.

\subsubsection{HTS algorithm}

Similarly to the case of nodes with hidden information, at the first stage of the algorithm, for each cascade $c$ we need to find the activation times which maximize the likelihood given the intervals they belong to and our current estimation of transmission probabilities:
\begin{equation}
\{\widehat{\tau^{c}_{i}}\}_{i \in V}=\arg\max_{\tau_i\in \left] (k^{c}_{i}-1) \Delta, k^{c}_{i} \Delta \right]} P(\{\tau^{c}_{i} \}_{i \in V}\mid {\widehat{G}_{\alpha}}).
\label{eq:likelihood_HTS_time}
\end{equation}

Again, we choose to approximately solve this optimization problem with a MC sampling of $L_{\Delta}$ cascades, using as sources $\{ i \mid i \in  V, \tau^{c}_{i} = 0 \}$. If the sample maximizing \eqref{eq:likelihood_HTS_time} contains activation times outside of the allowed interval, we fix them to the values found at a previous time step. The algorithm is initialized with $\tau^{c}_{i} = k^{c}_{i} \Delta$, and on the networks of size $N=20$ we used $L_{\Delta}=1000$ with $\Delta=2$ and $L_{\Delta}=10000$ with $\Delta=3$ and $\Delta=4$, the stopping threshold $\delta_{HTS}=10^{-2}$. 
 
\subsubsection{DMP algorithm}

In the case of the incomplete information in time, we choose the following objective function for the DMP algorithm:
\begin{equation}
J''=\sum_{t \in \mathcal{T}_{\mathcal{O}}}J''(t)=\sum_{k=1}^{K-1}\sum_{i \in V} \frac{1}{2} \left[\Delta\widetilde{P}^{i}_{S}(t^{(k)})-\Delta P^{i}_{S}(t^{(k)})\right]^{2},
\label{eq:mismatch_modified}
\end{equation}
where $\Delta P^{i}_{S}(t^{(k)}) = P^{i}_{S}(t^{(k)}) - P^{i}_{S}(t^{(k+1)})$, and $\Delta\widetilde{P}^{i}_{S}(t^{(k)})$ is an analogical empirical quantity from the data. Note that for $\Delta=1$, $\Delta P^{i}_{S}(t) = m^{i}(t)$, and $J''=J$. In this scheme, the couplings are updated according to $\alpha^{(t+\Delta)}_{rs} \leftarrow \alpha^{(t)}_{rs} - \epsilon \frac{\partial J''(t)}{\partial \alpha_{rs}}$ for $t \in \mathcal{T}_{\mathcal{O}}$ and fixed learning rate $\epsilon$, where
\begin{equation}
-\frac{\partial J''(t)}{\partial \alpha_{rs}} = \sum_{i \in V}\left[\Delta\widetilde{P}^{i}_{S}(t^{(k)})-\Delta P^{i}_{S}(t^{(k)})\right]\left(\frac{\partial P^{i}_{S}(t^{(k)})}{\partial \alpha_{rs}} - \frac{\partial P^{i}_{S}(t^{(k+1)})}{\partial \alpha_{rs}} \right).
\end{equation}
The derivative of $P^{i}_{S}(t)$ reads:
\begin{equation}
\frac{\partial P^{i}_{S}(t)}{\partial \alpha_{rs}} =  P^{i}_{S}(0) \sum_{k \in \partial i} p^{k \rightarrow i}_{rs}(t) \prod_{l\in \partial i \backslash k}\theta^{l \rightarrow i}(t),
\end{equation}
where $p^{k \rightarrow i}_{rs}(t)$ obeys the same set of dynamic message-passing equations \eqref{eq:SIequations:p}-\eqref{eq:SIequations:q} as before.

\subsubsection{Numerical results}

\begin{figure*}[!ht]
\begin{center}
\includegraphics[width=0.48\columnwidth]{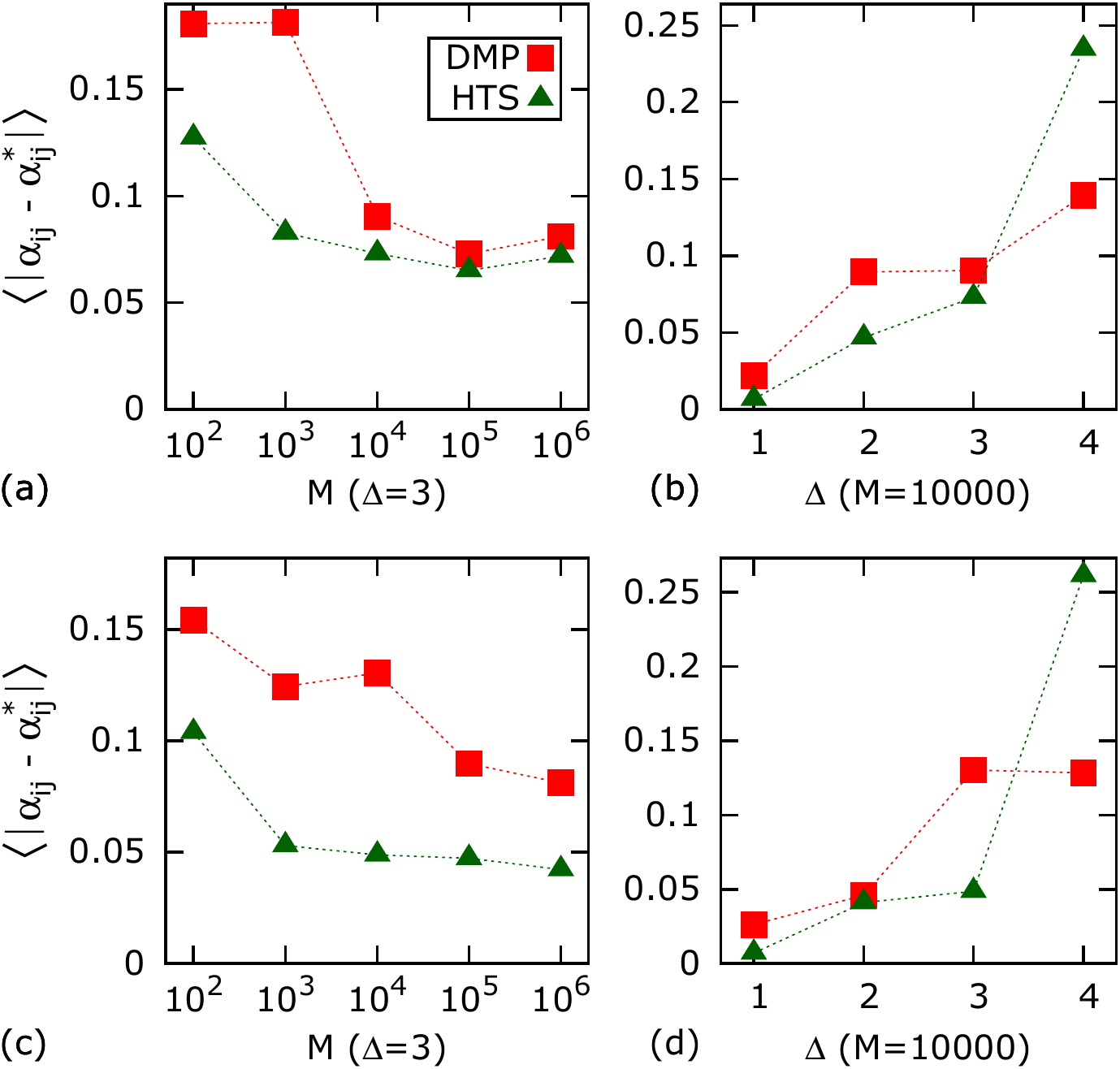}
\caption{(Color online) Test for the case of incomplete information in time. The results are presented for different values of parameters $M$ and $\Delta$ for (a), (b) a small tree network with $N=20$ nodes, and (c), (d) a connected component of a power-law network with $N=20$ and containing cycles of short length.}
\label{fig:Missing_Time}
\end{center}
\vspace{-5mm}
\end{figure*}   

The results for the case of missing information are presented in the Fig.~\ref{fig:Missing_Time} for a tree network (a)-(b) and for a connected component of a power-law network (c)-(d) with $N=20$ nodes. We chose uniform initial conditions $\alpha_{ij}=0.5$ $\forall (ij) \in E$ for the DMP algorithm in this case, although one could try to combine it with a ML starting point, as we have done for the case of the nodes with missing information. Again, HTS demonstrates a clearly better performance, but at an expense of a very large computational time (of order of a day for $\Delta=3$ and $M=10^4$) with respect to the DMP algorithm, which converges in less than a minute for such a small network.

Obviously, in the case of mixed partial observations (including nodes with hidden information and missing times), we simply have to eliminate the hidden nodes in the corresponding sum in the equation \eqref{eq:mismatch_modified}; we do not present results in this case in order to clearly draw a distinction between the effect of two situations. Finally, let us note that the distribution of observation times does not need to be uniform: if each observation is costly, one could optimize over the best subset of measurement times still requiring a reasonable reconstruction quality, an idea often used in the field called active learning.  

\subsection{Results for the real-world networks}

\begin{figure*}[!h]
\begin{center}
\includegraphics[width=0.5\columnwidth]{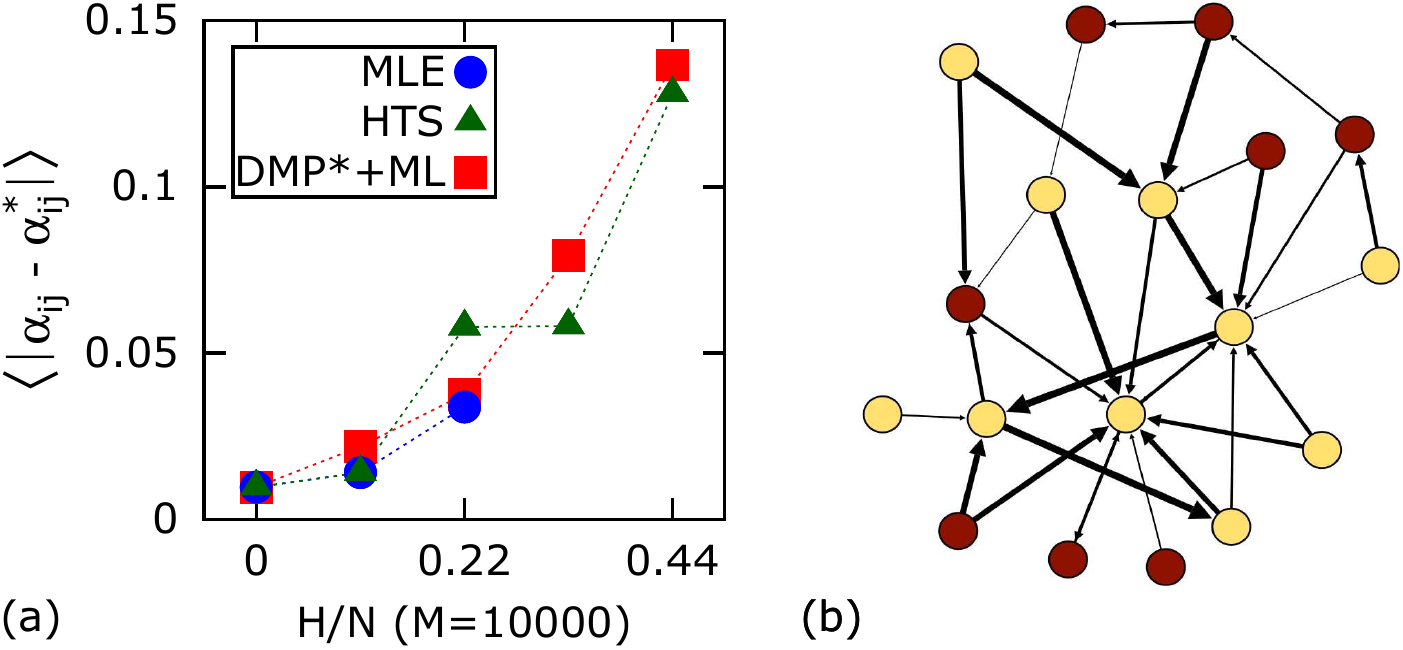}
\caption{(Color online) (a) Comparison of reconstruction performance of MLE, DMP*+ML and HTS algorithms for a real-world Sampson's monastery network with $N=18$ nodes \cite{sampson1969crisis,Pajek} for different numbers of hidden nodes $H$ but fixed number of cascades $M=10^4$. (b) Visualization of topology of the network obtained with Gephi \cite{bastian2009gephi}; a darker color indicates the location of nodes for which the information is missing.}
\label{fig:MON_M4}
\end{center}
\vspace{-5mm}
\end{figure*}

\begin{figure*}[!h]
\begin{center}
\includegraphics[width=0.3\columnwidth]{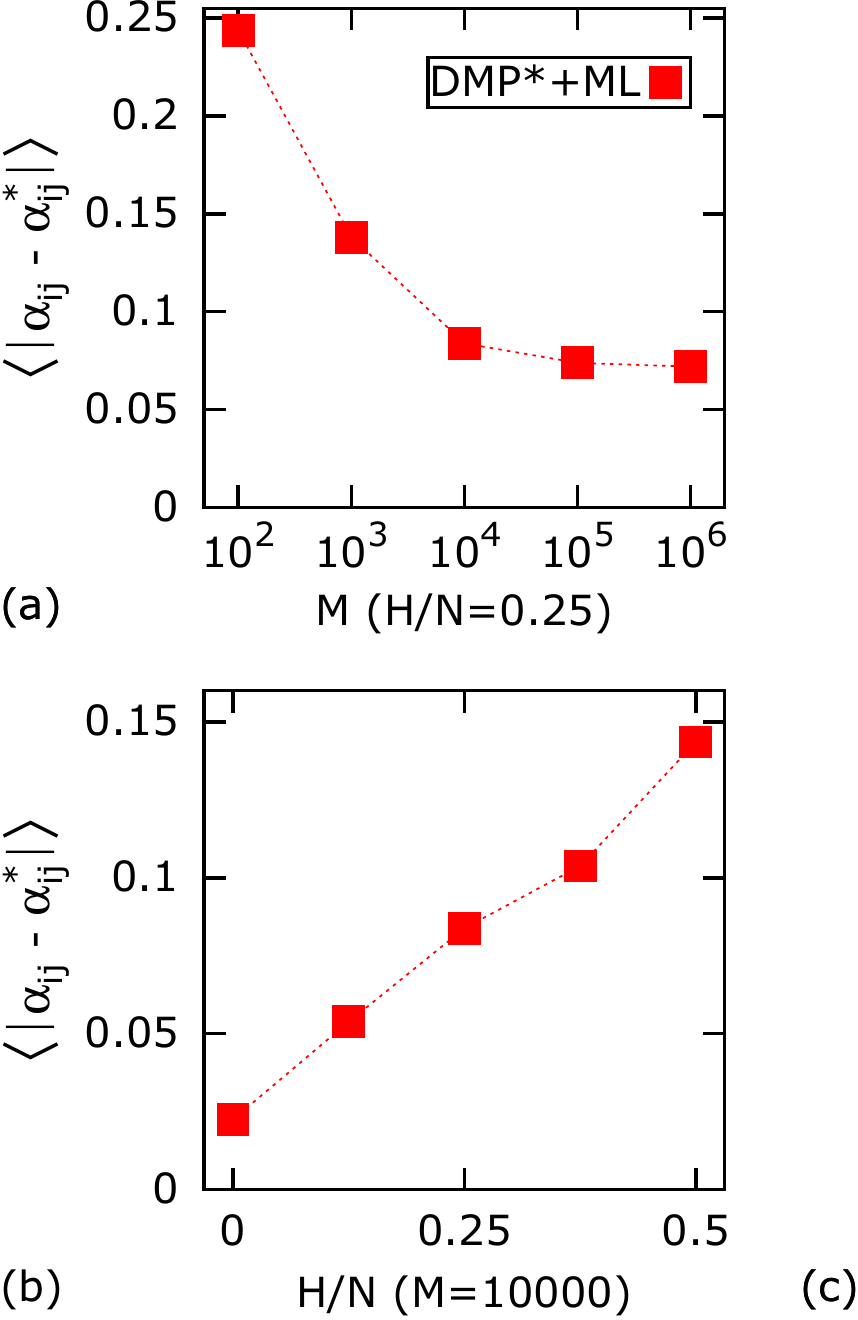}
\hspace{-0.1cm}\includegraphics[width=0.48\columnwidth]{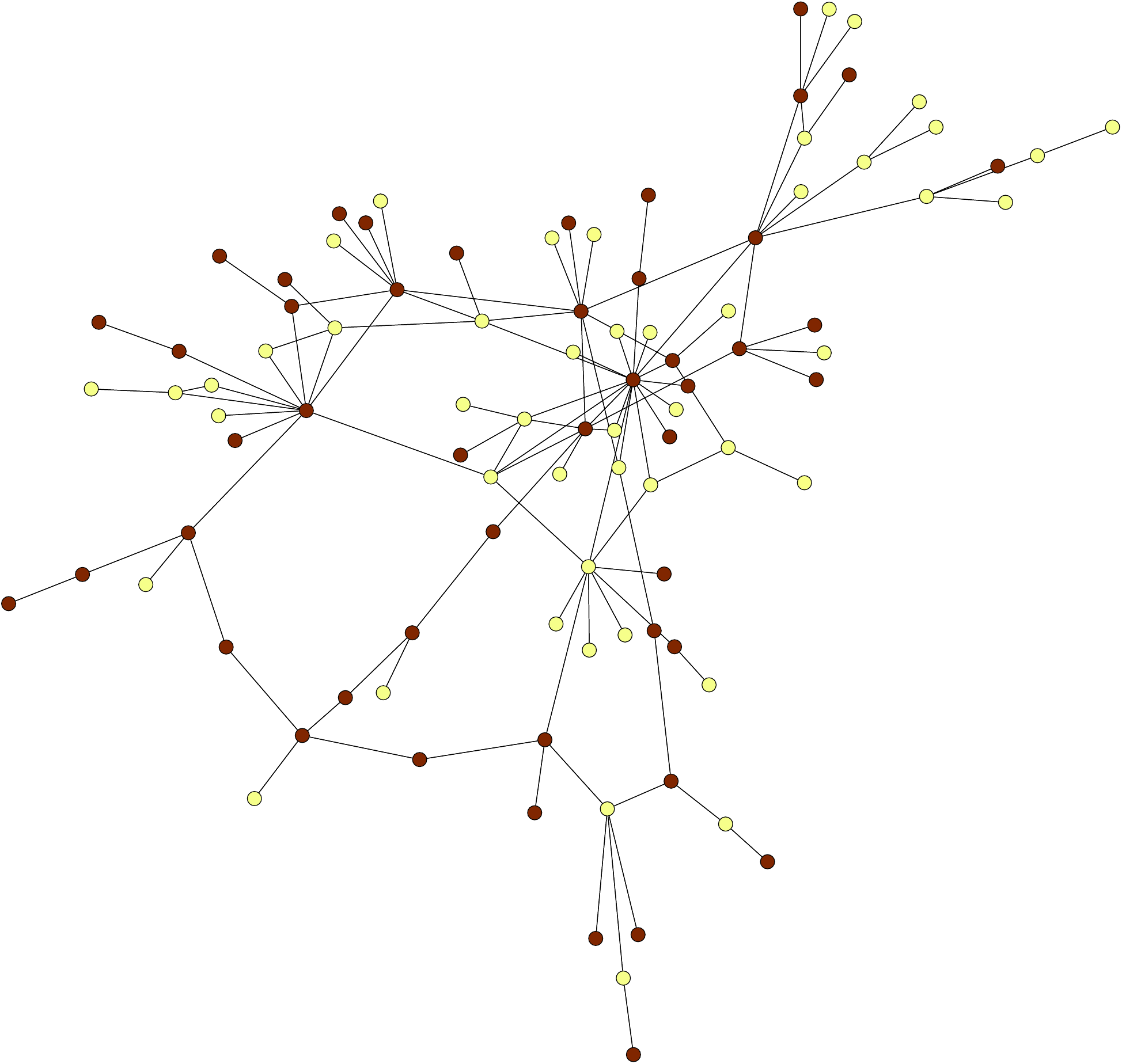}
\caption{(Color online) (a), (b) Reconstruction performance of the DMP*+ML algorithm on a network of retweets with $N=96$ nodes \cite{graphrepository2013}. (c) Topology of the social network with a darker color indicating the location of $48$ (half of the total number of vertices) randomly chosen hidden nodes.}
\label{fig:Twitter}
\end{center}
\vspace{-5mm}
\end{figure*}

In this section, we present some supplementary numerical studies of reconstruction algorithms for different real-world networks. In the Fig.~\ref{fig:MON_M4}, we represent an average reconstruction error on the Sampson's monastery network \cite{sampson1969crisis,Pajek} for different $H$ (up to eight hidden nodes which locations are depicted in the Fig.~\ref{fig:MON_M4} (b)) with $M=10^4$ cascades. As highlighted in the main text, the DMP algorithm demonstrates a competitive quality of reconstruction with a major gain in the computational speed. Essentially, the algorithms based on the maximization of the incomplete likelihood are not applicable even to the networks of a moderate size, such as the network of retweets with $N=96$ nodes \cite{graphrepository2013}. Hence, only results for the DMP*+ML algorithm are presented for this example of a real social network, cf. Fig.~\ref{fig:Twitter}; the outcome for this network is quite reasonable, with an accuracy comparable to the one obtained in the study of the Sampson's monastery network.       

Let us nonetheless also provide an example of a dense topology on which both algorithms fail, the DMP-based one being especially inefficient. An instance of such a topology is given by the network of political alliances and enmities among the 16 tribes of Eastern Central Highlands of New Guinea \cite{Pajek}, documented by Read \cite{read1954cultures}. Because of the high density of the network, the hidden nodes introduce a high degeneracy of the possible solution and play a more important role compared to the case of sparse networks, so that when the fraction of hidden nodes is of order of a half of the network, the reconstruction algorithms do not yield any useful information on the couplings (the demonstrated error is of the same order as the one at the initial point, where all transmission probabilities are set to $0.5$).  

\begin{figure*}[!h]
\begin{center}
\includegraphics[width=0.8\columnwidth]{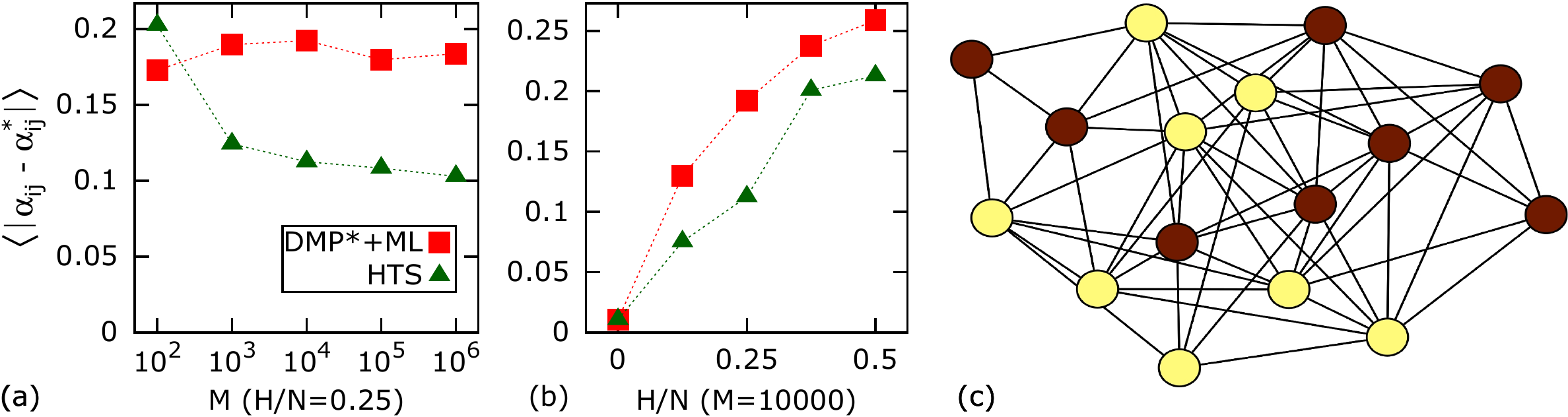}
\caption{(Color online) (a), (b) Reconstruction performance of DMP*+ML and HTS algorithms on a network representing the alliance structure among $N=16$ tribes of New Guinea \cite{read1954cultures}. (c) Network topology, visualized with Gephi \cite{bastian2009gephi}; a darker color indicates the location of nodes with hidden information.}
\label{fig:Tribes}
\end{center}
\vspace{-5mm}
\end{figure*}

\subsection{DMP algorithm for continuous dynamics}

All the results in this Letter have been presented for the discrete-time formulation insofar. In this section, we briefly discuss how the same techniques can be easily extended to continuous case. The maximum likelihood estimator has been originally suggested for the continuous dynamics of the independent-cascade model \cite{Gomez-Rodriguez2011}; our extension of the MLE for the case of missing information follows straightforwardly the main text, with discrete sums replaced by the integrals. For the DMP algorithm, we can use the continuous version of the DMP equations for the SI model, derived for the first time in \cite{KarrerNewman2010}. An important difference only concerns a choice of the objective function $J$: as in the case of missing information in time, it is more convenient to quantify the mismatch in terms of differences of probabilities $P^{i}_{S}(t)$ at a certain number of discrete times:
\begin{equation}
J''=\sum_{t=0}^{n}J''(t)=\sum_{t=0}^{n-1}\sum_{i \in \mathcal{O}} \frac{1}{2} \left[\Delta\widetilde{P}^{i}_{S}\left(\frac{t}{n}T\right)-\Delta P^{i}_{S}\left(\frac{t}{n}T\right)\right]^{2},
\end{equation}
where $n$ is the number of discretization steps in time which should be related to the statistics of activation times in $M$ observed cascades, or to the set of observation times in the case of incomplete information in time, while $\Delta\widetilde{P}^{i}_{S}(t)$ and $\Delta P^{i}_{S}(t)$ are defined in a way similar to the case of incomplete time information.

In the case of constant rates $\alpha_{ij}$, we define the transmission function as $f_{ij}(t) = \alpha_{ij} e^{- \alpha_{ij} t}$. Then the functions $\theta^{i \rightarrow j}(t)$ are computed as follows \cite{KarrerNewman2010}:
\begin{align}
\notag
\theta^{i \rightarrow j}(t) & = 1 - \int_0^t d \tau \ f_{ij}(\tau) \left[ 1 - P_S^i (0) \prod_{k \in \partial i \backslash j } \theta^{k \rightarrow i}(t - \tau)  \right]
\\
&= e^{- \alpha_{ij} t} +  P_S^i (0) \alpha_{ij} e^{- \alpha_{ij} t} \int_0^t d \tau \  e^{ \alpha_{ij} \tau} \left( \prod_{k \in \partial i \backslash j} \theta^{k \rightarrow i}( \tau) \right).
\label{eq:continous_integral}
\end{align}
In order to compute the dynamic messages $\theta^{i \rightarrow j}(t)$, we can either integrate the expression \eqref{eq:continous_integral} numerically, or transform the equation above into an ordinary differential equation by integrating the last term in \eqref{eq:continous_integral} by parts:
\begin{equation}
\frac{d \theta^{i \rightarrow j}(t)}{dt} =  - \alpha_{ij} \theta^{i \rightarrow j}(t) +  \alpha_{ij} P_S^i (0) \prod_{k \in \partial i \backslash j} \theta^{k \rightarrow i}(t),
\label{eq:continous_differential}
\end{equation} 
which can be solved numerically starting from initial conditions $\theta^{i \rightarrow j}(0)=1$. The probabilities $P^{i}_{S}(t)$ are computed according to \eqref{eq:SIequations:P_S} in the continuous case as well.

The couplings are updated according to $\alpha^{(t+\Delta t)}_{rs} \leftarrow \alpha^{(t)}_{rs} - \epsilon \frac{\partial J''(t)}{\partial \alpha_{rs}}$ for $t \in [0,n]$, $\Delta t = T/n$, and fixed learning rate $\epsilon$, where
\begin{equation}
-\frac{\partial J''(t)}{\partial \alpha_{rs}} = \sum_{i \in \mathcal{O}}\left[\Delta\widetilde{P}^{i}_{S}\left(\frac{t}{n}T\right)-\Delta P^{i}_{S}\left(\frac{t}{n}T\right)\right]\left(\frac{\partial P^{i}_{S}\left(Tt/n\right)}{\partial \alpha_{rs}} - \frac{\partial P^{i}_{S}\left(T(t+1)/n\right)}{\partial \alpha_{rs}}\right).
\end{equation}
The derivative of $P^{i}_{S}\left(Tt/n\right)$ reads:
\begin{equation}
\frac{\partial P^{i}_{S}\left(Tt/n\right)}{\partial \alpha_{rs}} =  P^{i}_{S}(0) \sum_{k \in \partial i} p^{k \rightarrow i}_{rs}(Tt/n) \prod_{l\in \partial i \backslash k}\theta^{l \rightarrow i}(Tt/n),
\end{equation}
where $p^{k \rightarrow i}_{rs}(t)$ obeys the following ordinary differential equation, obtained by taking the derivative of \eqref{eq:continous_differential}:
\begin{align}
\notag
\frac{d p^{k \rightarrow i}_{rs}(t)}{d t} = - \alpha_{ki} p^{k \rightarrow i}_{rs}(t) &+ \alpha_{ki} P^{k}_{S}(0) \sum_{m \in k \backslash i} p^{m \rightarrow k}_{rs}(t) \prod_{l\in \partial k \backslash \{i,m\}}\theta^{l \rightarrow k}(t)
\\
& + \mathds{1}[k=r,i=s] \left[ - \theta^{k \rightarrow i}(t) +  P_{S}^{k}(0) \prod_{l \in \partial k \backslash i} \theta^{l \rightarrow k}(t) \right]. 
\end{align}

\subsection{Perspectives of the structural learning using the DMP algorithm}

\begin{figure*}[!ht]
\begin{center}
\includegraphics[width=0.88\columnwidth]{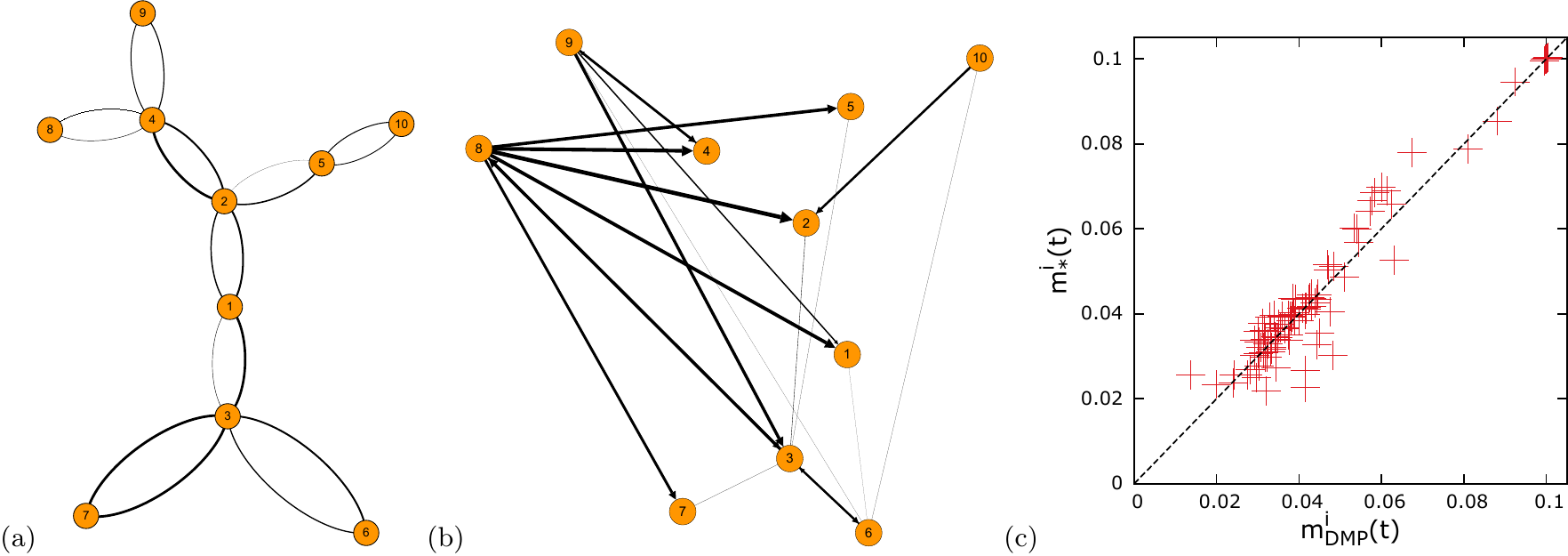}
\caption{(Color online) Results given by the DMP reconstruction algorithm in the case where the underlying network is unknown: (a) Topology of the original network (a tree with $N=10$ nodes), (b) Topology of the reconstructed network, (c) Scatter plot of marginal probabilities $m^{i}(t)$: empirical from the simulations $m^{i}_{*}(t)$ versus those predicted by DMP $m^{i}_{\text{DMP}}(t)$. Thickness of network edges are proportional to the (a) true and (b) estimated transmission probabilities, the direction of the edges on the graph (a) should be read clockwise from source to target. Both networks are visualized with Gephi \cite{bastian2009gephi}.}
\label{fig:FCG_reconstruction}
\end{center}
\vspace{-5mm}
\end{figure*}

In this section we show that the DMP algorithm in its simplest form is not suitable for a reconstruction of the network structure. As an example, we simulated $M=10^6$ cascades for $T=10$ time steps on a small tree network with $N=10$ nodes; its topology is depicted in the Fig.~\ref{fig:FCG_reconstruction}~(a). Precomputed empirical marginal probabilities have been given to the DMP algorithm, and since the underlying network is supposed to be unknown, we initialized the algorithm on a fully-connected graph with all couplings equal to $0.5$. As demonstrated in the Fig.~\ref{fig:FCG_reconstruction}~(b), the resulting DMP-reconstructed network is sparse, but does not have the exact topology of the original graph. Notice that the DMP algorithm does exactly what it is supposed to do: it tries to match the marginal probabilities given by the DMP equations with the empirical marginal probabilities from the simulations, see Fig.~\ref{fig:FCG_reconstruction}~(c). It turns out that the algorithm sets the majority of transmission probabilities to zero, reconstructing a sparse directed network that fits in the best way to the input data. In order to get an idea about the similarity of the original and the DMP-reconstructed networks, we summarize some basic topological parameters of both graphs in the Table~\ref{table:topological_parameters}.

We saw that because of the loss of information due to averaging of original data, the DMP algorithm in its simplest form does not lead to a correct reconstruction of the diffusion network, even in the case of fully available observations. Let us however outline some situations in which the use of DMP for structural learning would most probably make sense, especially in the presence of hidden nodes. In this case, all the algorithms would require some additional assumptions on the network structure. If a part of the network is known, this would likely impose an additional constraint on the DMP algorithm, forcing it to reproduce the original topology. If no information on the location of the hidden node is available, which would be a hard case for all reconstruction algorithms, one could think of the use of DMP algorithm with $\ell_{1}$ regularization to speed up the learning process in the same way as we used the DMP algorithm for parameter learning in this Letter. We demonstrated that parameter learning in the presence of missing information is already a hard problem even when the underlying network is known; we anticipate that the structural learning with latent variables and without any prior information on the network would be even harder from the algorithmic point of view. The most direct way to solve this problem would probably be based on the information contained in the effective interactions at several time steps created via a marginalization over hidden nodes, and using correlations between individual cascades in order to distinguish between different latent nodes. These interesting questions are left for the future work. 

However, let us note that the DMP algorithm in its present form could be straightforwardly used for the network engineering purposes, i.e. for building a sparse network with some prescribed properties with respect to a given cascading process given, for instance, by the average probabilities of activation at certain times.   

\begin{center}
\begin{table}[!hb]
\begin{tabular}{| c || c | c | c |}
\hline
	& Original & Reconstructed
\\ 
\hline
\hline
average degree & 1.8 & 2.2\\
average weighted degree & 1.685 & 0.583\\
graph density & 0.200 & 0.244\\ 
diameter & 5 & 5\\
number of shortest paths & 90 & 72\\
average path length & 2.82 & 2.28\\
\hline
\end{tabular}
\caption{Comparison of topological parameters between original and DMP-reconstructed graphs.}
\label{table:topological_parameters}
\end{table}
\end{center}

\end{document}